\documentclass[aps,floatfix,nofootinbib,twocolumn,prl]{revtex4-1}

\usepackage{graphicx}
\usepackage{epic}
\usepackage{eepic}
\usepackage{latexsym}

\newcommand{\eq}[1]{(\ref{#1})}
\newcommand{\be}{\begin{equation}}
\newcommand{\ee}{\end{equation}}
\newcommand{\bea}{\begin{eqnarray}}
\newcommand{\eea}{\end{eqnarray}}

\newcommand{\hs}[1]{\hspace{#1 mm}}

\newcommand{\cH}{{\cal H}}

\def\C{\Gamma}
\def\d{\delta}

\def\e{\epsilon}

\def\f{\phi}
\def\fr{\frac}

\def\k{\kappa}
\def\l{\lambda}

\def\m{\mu}
\def\n{\nu}

\def\r{\rho}
\def\s{\sigma}
\def\S{\Sigma}

\def\O{\Omega}

\def\o{\omega}

\def\del{\partial}

\let\bm=\bibitem
\def\nn{\nonumber}

\begin{document}
\large

\title{ \Large Comments on the Canonical Measure in Cosmology }

\author{Ali Kaya}
\email[]{ali.kaya@boun.edu.tr}
\affiliation{\large Bo\~{g}azi\c{c}i University, Department of Physics, \\ 34342,
Bebek, Istanbul, Turkey }

\begin{abstract}
In the mini-superspace approximation to cosmology, the canonical measure can be used to compute probabilities when a cutoff is introduced in the phase space to regularize the divergent measure. 
However, the region initially constrained by a simple cutoff evolves non-trivially under the Hamiltonian flow. 
We determine the deformation of the regularized phase space along the orbits when a cutoff is introduced for the scale factor of the universe or for the Hubble parameter. In the former case, we find that the cutoff for the scale factor varies in the phase space and effectively decreases as one evolves backwards in time. In the later case, we calculate the probability of slow-roll inflation in a chaotic model with a massive scalar, which turns out to be cutoff dependent but not exponentially suppressed. We also investigate the measure problem for non-abelian gauge fields giving rise  to inflation. 
\end{abstract}

\maketitle

\section{Introduction}

In an interesting paper \cite{gt}, Gibbons and Turok claimed that the probability of getting  $N$ e-folds of inflation is  exponentially suppressed by  a factor of $\exp(-3N)$. They used the canonical measure on the constrained phase space in the mini-superspace approximation \cite{han,ghs} to define a flat probability distribution for solutions. It is known that the canonical measure diverges  and thus it is difficult to determine the probability of inflation even in the mini-superspace approximation \cite{hp} (the situation is the same for $R^2$  inflation \cite{p1} and for the anisotropic  Bianchi type-I model \cite{p2}). In \cite{gt}, by arguing that universes larger than a critical size cannot be observationally distinguished, Gibbons and Turok introduced a cutoff for the scale factor of the universe to make the measure finite (see e.g. \cite{lw} and \cite{p3} for generalization of their method to different cases and see also \cite{rb} for an alternative approach to the measure problem).  

A nice feature of this construction is that Liouville's theorem guarantees "time-independence" of the assigned probabilities. Actually, in this context one needs a theorem which is slightly different than the standard Liouville's theorem since orbits in cosmology are necessarily constrained in the reduced phase space with vanishing Hamiltonian. In \cite{ghs} a theorem of that sort is proved, which essentially states that in a proper measure any hypersurface {\it transversely} intersecting the orbits in the reduced phase space can be used to calculate probabilities. Using this invariance, in \cite{gt} Gibbons and Turok choose  a surface of constant Hubble parameter $H$, which is a suitable surface since $H$ is monotonic for the flat and the hyperbolic Freedman-Robertson-Walker (FRW) models. For regularization,  they introduce a cutoff for the scale factor of the universe at low enough $H$ such that the evolution of the scalar field is adiabatic. 

In this paper, we note that a {\it simple} region restricted by a cutoff as in \cite{gt} may deform non-trivially when it is slid along the orbits. Specifically, we show that the cutoff for the scale factor {\it effectively} decreases as the surface evolves backwards in time. Moreover, the cutoff becomes field dependent due to different  amount of shifts along the orbits. Therefore, one ends up a varying cutoff across the phase space, which would impair the naturalness of the procedure. 

In this work, we also consider surfaces of constant scale factor $a=a_*$ as the transverse initial value surfaces in the flat FRW model and introduce a cutoff for the Hubble parameter to make the probability measure finite. This foliation has already been discussed in earlier works, see \cite{hp,c}. Note that a cutoff in $H$ (presumably near the Planck scale) is already required  for the validity of the classical field equations. In this case the probability of slow-roll inflation  in a chaotic model with a massive scalar turns out to be cutoff dependent but not exponentially  suppressed. In principle, by the theorem of \cite{ghs} the assigned probabilities should not change for different transverse hypersurfaces (e.g. constant $a$ or constant $H$ surfaces). However, the regions regulated by the two different cutoffs become very distinct from each other giving two different results. 

Finally, in this paper we study the measure problem for non-abelian gauge fields giving rise to inflation \cite{msj}. Apart from minor peculiarities, the situation with the gauge fields turns out to be  similar to the scalars, namely the measure diverges and different cutoffs can be introduced for regularization.  In the conclusions, we discuss the implications of these findings for the measure problem in cosmology. 

 \section{The canonical measure and Liouville's theorem}
 
Consider a finite dimensional  dynamical system governed by a Hamiltonian $\cH$. In the $2n$-dimensional phase space $\C$,  the symplectic form may be written in the local Darboux coordinates as
 \be
 \o=\sum_{i=1}^n dp_i\wedge dq_i.
 \ee
A natural volume element in the phase space can be obtained from the symplectic 2-form by wedge product, which gives $\O=\o\wedge...\wedge\o\equiv\o^n$. Liouville's theorem can be proved by noting that  $\o$, and thus $\O$, is invariant under the Hamiltonian flow
\be
{\cal L}_{X_\cH} \o=0,
 \ee
where ${\cal L}$ denotes the Lie derivative and $X_\cH$ is the  Hamiltonian vector field which is defined by $\o(X_\cH)=d\cH$.  In cosmology, however, the Hamiltonian vanishes and the system is necessarily constrained in the $(2n-1)$-dimensional subspace $\tilde{\C}$ defined by $\cH=0$.  Therefore, in its standard form Liouville's theorem is not applicable. In \cite{ghs}, a similar theorem suitable for cosmology in the mini-superspace approximation is proved as follows: If  $\cH$ is chosen as one of the momentum coordinates $P=\cH$, then the symplectic form can be written as
\be
\o=dP\wedge dQ+\o_\cH,
\ee
where $Q$ is the coordinate conjugate to $P$ and $\o_\cH$ is a closed two-form $d\o_\cH=0$, which can be seen as a symplectic structure of a $(2n-2)$-dimensional space. Note that $\o_\cH=\o|_{\tilde{\C}}$, i.e. $\o_\cH$ can  be obtained from the restriction of $\o$ on $\tilde{\C}$. In these coordinates, the Hamiltonian vector field becomes $X_\cH=\del/\del Q$ and the equations of motion implies $Q=t$.  It is  now easy to see that for an {\it arbitrary} function $f$,
\be\label{l}
{\cal L}_{fX_\cH} \o_\cH=d(fX_\cH.\o_\cH)+fX_\cH.(d\o_\cH)=0,
\ee
where the dot denotes contraction of a differential form by a vector field. Let $\S$ be a $(2n-2)$-dimensional space transverse to the Hamiltonian flow in the  constrained phase space $\tilde{\C}$. Then, \eq{l} implies  
\be
\int_\S  \o_\cH^{(n-1)}= \int_{\S'} \o_\cH^{(n-1)},
\ee
where $\S'$ is another surface transverse to the Hamiltonian flow, which can be obtained from $\S$ by sliding along the orbits of $fX_\cH$.  

Let us illustrate this theorem for a simple system, also studied in \cite{ghs}, which will be useful for our discussion in the next section. Consider a free particle moving in two dimensions, which has the Hamiltonian 
\be
\cH=\fr12 (p_x^2+p_y^2).
\ee
The constrained phase space $\cH=E$ can be parametrized by the coordinates $(x,y,p_x)$ and  $p_y$ can be solved as 
\be\label{y}
p_y=+\sqrt{2E-p_x^2},
\ee
where  we simply select orbits with increasing $y$.  From the symplectic form $\o=dp_x\wedge dx+dp_y\wedge dy$,  one can find 
\be
\o_\cH=-\fr{p_x}{\sqrt{2E-p_x^2}}dp_x\wedge dy+dp_x\wedge dx.
\ee
It is easy to check that 
\be
{\cal L}_{fX_\cH} \o_\cH=0,
\ee
where $X_\cH=p_x\del_{x}+p_y\del_{y}$. 

\begin{figure}
\leftline{\hs{10}\includegraphics[width=6.8cm]{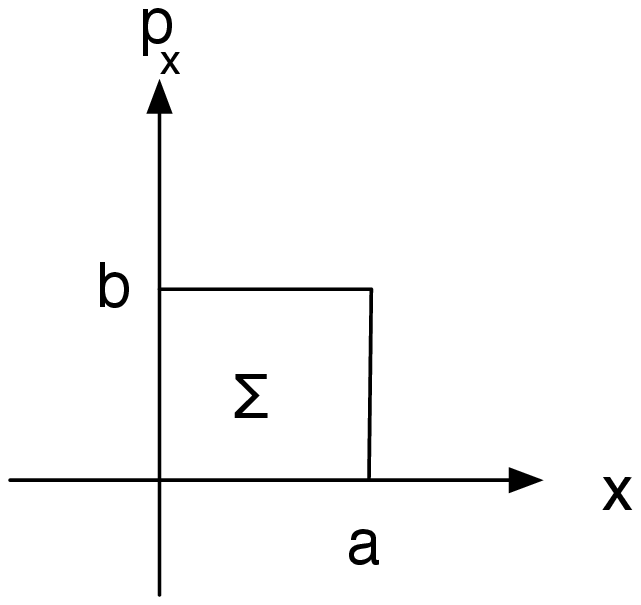}}
(a)
\leftline{\hs{10}\includegraphics[width=4.5cm]{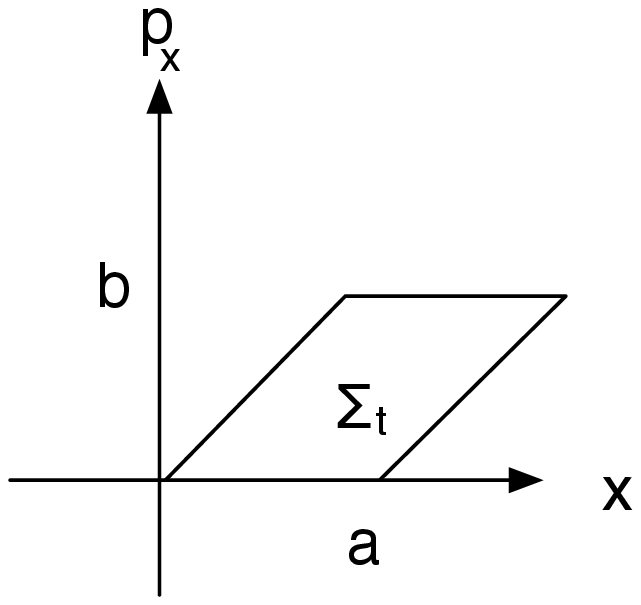}}
(b)
\caption{(a) The region $\S$ at $t=0$. (b) Evolution of $\S$ under the Hamiltonian flow with $f=1/p_y$.} 
\label{fig1}
\end{figure}

The surface $\S:y=c$  is transverse to  the orbits\footnote{Orbits with $p_y=0$ do not intersect  $\S$, but they are of measure zero in the phase space.} and  $\o_\cH$ can be integrated over it. However, the integral diverges since
\be
\int_\S \o_\cH=\int_{-\sqrt{2E}}^{\sqrt{2E}}\int_{-\infty}^\infty dx \,dp_x\to\infty. 
\ee
Let us, therefore, define a restricted region  by $\S:y=c;\,0<x<a;\,0<p_x<b$, which gives
\be
\int_\S \o_\cH=\int_0^{b}\int_{0}^a dx \,dp_x=ab. 
\ee 
Now, it is easy to see that the region $\S':y=c';\,0<x<a;\,0<p_x<b$ cannot be obtained from $\S$ by sliding along the orbits. Instead, if one chooses $f=1$ then the surface slid along the orbits by a parameter $t$ becomes 
\bea
&&\S_t:y=c+p_y t;\nn\\
 &&0<p_x<b; \,\,\,p_x t<x<a+p_x t,\nn
\eea
where as noted above $p_y$ is seen as a function of $p_x$ given by \eq{y}.  A nicer choice is $f=1/p_y$ which gives a surface in the $(x,p_x)$ plane given by   (see Fig. \ref{fig1})
\bea\nn
&&\S_t:y=c+ t;\\
&& 0<p_x<b; \,\,\,(p_x/p_y) t<x<a+(p_x/p_y) t.\nn
\eea
In either case one can check that $\int_{\S_t} \o_\cH=ab$, as it must be. 
 
This simple example shows that in applying the  theorem of \cite{ghs} to a restricted region, one must take into account the evolution along the orbits carefully, which can be non-trivial. Namely, a seemingly "ordinary" region may deform along the orbits as in Fig. \ref{fig1}. 
 
\section{The measure in cosmology}

Consider a cosmological model with a massive scalar field in the mini-superspace approximation where the action
\be\label{ac}
S=\fr12\int\sqrt{-g}\left[R-(\nabla\phi)^2-m^2\phi^2\right]
\ee
reduces to 
\be\label{rac} 
S=\int\, dt \left[-\fr{3a}{N}\dot{a}^2+\fr{a^3}{2N}\dot{\f}^2-\fr{Na^3}{2}m^2\f^2\right],
\ee
after assuming $\f=\f(t)$ and 
\be\label{met}
ds^2=-N(t)^2dt^2+a(t)^2(dx^2+dy^2+dz^2).
\ee
Although it is possible to work out flat ($k=0$), hyperbolic ($k=-1$) and spherical ($k=1$) FRW models  simultaneously, here we consider the flat model, which is geometrically more transparent. Note that models with different values of $k$  must be distinguished from each other,  i.e. the constant spatial curvature of space is chosen from the beginning as a parameter of the model, and once it is fixed the dynamical evolution cannot change it. 

From \eq{rac} it is straightforward to determine the canonical momenta 
\be\label{mom}
p_a=-\fr{6a\dot{a}}{N},\hs{7}p_\f=\fr{a^3\dot{\f}}{N},
\ee
and the Hamiltonian 
\be\label{ham}
\cH=N\left[-\fr{1}{12}p_a^2+\fr{1}{2a^3}p_\f^2+\fr{a^3}{2}m^2\f^2\right].
\ee
Varying  Hamiltonian with respect to the lapse function $N$ gives the constraint after which one can set $N=1$. For calculations  it is convenient to use non-canonical coordinates $(H,a,\dot{\f},\f)$ instead of $(p_a,a,p_\f,\f)$, where $H$ is the Hubble parameter $H=\dot{a}/a$. Note that $a\in R^+$ since $a=0$ is a singular point, e.g. the Hamiltonian \eq{ham} became undefined. In these coordinates the Hamiltonian vector field becomes
\be\label{hv}
X_\cH=aH\del_a+\dot{\f}\del_\f-\left[3H\dot{\f}+m^2\f\right]\del_{\dot{\f}}-\fr12\dot{\f}^2\del_H.
\ee
Fixed points of the Hamiltonian flow are given by  $H=0$, $\f=0$ and $\dot{\f}=0$, which correspond to the flat space with different "radii" $a$. 

One may tempt to identify two space-times which have the same $H$, $\f$ and $\dot{\f}$ values but different scale factors $a$, since  a coordinate change $(x,y,z)\to\l(x,y,z)$ implies $a\to\l a$ (this was identified in \cite{as} as an extra gauge symmetry of the flat model). This freedom can easily be remedied by taking $(x,y,z)$ to be 3-torus making $a$ to be the radius (this compactification is  indeed necessary for a proper reduction of the action \eq{ac} to \eq{rac}). Even for the non-compact case, one can  set a unit system for $(x,y,z)$ and thus prohibit the free scaling of these coordinates. Therefore, space-times with different scale factors can be thought to be physically distinct. 

\begin{figure}
\centerline{\includegraphics[width=6.8cm]{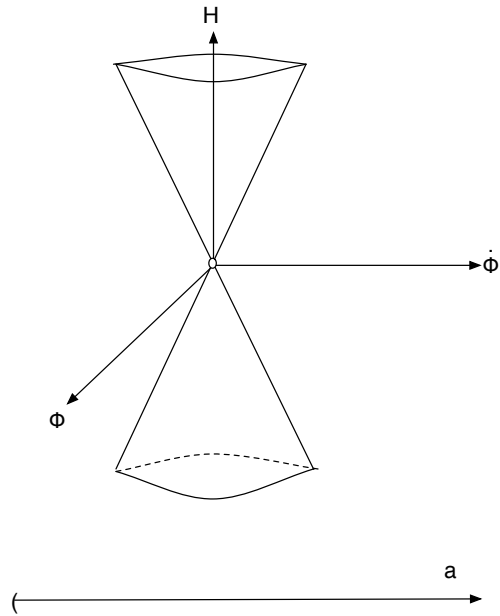}}
\caption{The constrained phase space, which is given by $C_2\times R^+$.} 
\label{fig2}
\end{figure}

In the non-canonical coordinates, the constrained phase space $\tilde{\C}$, defined by $\cH=0$, becomes
\be\label{c}
6H^2-\dot{\f}^2- m^2\f^2=0.
\ee
Therefore $\tilde{\C}$ is given by $C_2\times R^+$, where $C_2$ is the two-dimensional cone in $(H,\f,\dot{\f})$ space defined by \eq{c} and $R^+$ stands for the scale factor $a$ (see Fig. \ref{fig2}). If one removes the origin, which is the fixed point of the flow, then the  phase space for the expanding solutions $H>0$  becomes $S^1\times R^+\times R^+$. Note that expanding and the contracting orbits are dynamically separated from each other by the fixed point.  

The symplectic form $\o=dp_a\wedge da+dp_\f\wedge d\f$ can uniquely be reduced to the constrained phase space to give $ \o_{\cH}$, which can be expressed in three equivalent ways by solving either $H$, $\f$ or $\dot{\f}$ from \eq{c} in terms the other two coordinates. One then needs to identify a surface $\S$ transverse to the orbits in $\tilde{\C}$ to define the canonical probability distribution. Since $H$ and $a$ are monotonic in the flat FRW model, there are two main ways of introducing such a surface. 

\subsection{Surfaces of constant  $H$}

In \cite{gt}, $\S$ is chosen to be the surface defined by $H=H_*$ in $\tilde{\C}$. Topologically it is given by $\S=S^1\times R^+$, where $S^1$ is the circle in the $(\f,m\dot{\f})$ plane defined by \eq{c} with $H=H_*$ and $a\in R^+$. Note that the orbits passing through $\dot{\f}=0$ are actually tangential to $\S$ but this does not cause a problem since such orbits form a set of measure zero in $\S$.  Following \cite{gt}, it is convenient to choose $(\f,a)$  as coordinates in $\S$, which gives
\bea
\o_\cH&=&dp_\f \wedge d\f\nn\\
&=&3a^2\sqrt{6H_*^2-m^2\f^2}\,da\wedge d\f.\label{int}
\eea
However, $\int_\S \o_\cH$ diverges as $a\to\infty$, therefore it is not possible to use this measure to define a probability distribution in the solution space \cite{hp}. 

\begin{figure}
\centerline{\includegraphics[width=8cm]{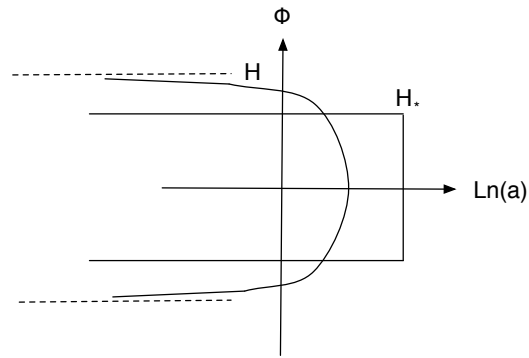}}
\caption{The surface $\S$, which is simply restricted by a cutoff at $H_*$ by $a<a_c$, deforms non-trivially at a larger Hubble parameter H. The deformed surface intersects the $\f=\pm\sqrt{6}H/m$ lines at infinity.}
\label{fig3}
\end{figure}

In \cite{gt}, it is proposed to introduce a cutoff for the scale factor to make the measure finite. The restricted surface transverse to the orbits  is chosen to be $\S=S^1\times I$, where the finite interval $I$, which replaces $R^+$, is given by $I: 0<a<a_c$. On $\S$, the measure can be evaluated  as\footnote{The integral \eq{i1} is actually over $S^1$, therefore after integrating $m\f$ from $-\sqrt{6}H_*$ to $+\sqrt{6}H_*$ the result must be multiplied by 2.}
\be\label{i1}
\int_\S \o_\cH=\fr{6\pi}{m}a_c^3 H_*^2. 
\ee
The theorem of \cite{ghs} guarantees that the value of this integral does not change under smooth deformations of $\S$ along the orbits. One may then think that the when the surface deformed to a different Hubble parameter $H$,  a new cutoff $a_c(H)$ must be defined by 
\be\label{cutoff}
a_c(H)=\left(\fr{H_*}{H}\right)^{2/3}\,a_c,
\ee
such that \eq{i1} does not change. But, as noted in the previous section the evolution of $\S$ along the orbits can be quite complicated, i.e. the "shape" may deform in a non-trivial  way (see Fig \ref{fig1}). Therefore, let us try to determine the evolution of $\S$ under a general flow $fX_\cH$. From \eq{hv} one may find that
\bea
&&\fr{d\f}{dt}=\pm f\,\sqrt{6H^2-m^2\f^2},\\
&&\fr{dH}{dt}=f\left(\fr12 m^2\f^2-3H^2\right),\label{ht}\\
&&\fr{d\dot{\f}}{dt}=-f\left(3H\dot{\f}+m^2\f\right),\\
&&\fr{da}{dt}=faH.\label{at}
\eea
Using $H$ as the "time" parameter gives\footnote{The singularity at $\f=\pm\sqrt{6}H/m$ is due to breakdown of $H$ as a nice flow parameter by \eq{ht}.}
\bea
&&\fr{d\f}{dH}=\mp\fr{2}{\sqrt{6H^2-m^2\f^2}},\label{a0}\\
&&\fr{da}{dH}=-\fr{2aH}{6H^2-m^2\f^2}. \label{ad}
\eea
Thus, flow equations in $H$ become independent of the arbitrary function $f$. Although an analytical solution is difficult to obtain, it is now possible to use \eq{a0} and \eq{ad} together with the initial conditions $a\leq a_c$ and $-\sqrt{6}H_*/m<\f<\sqrt{6}H_*/m$ at $H=H_*$ to find the new surface $\S_H$ at time $H$. Specifically,  \eq{ad} shows that the sliding of the scale factor along $H$ depends on the value of the scalar field and thus the new surface\footnote{Of course, it is possible to consider different deformations of $\S$ along the orbits. For example, if one chooses $f=1/(aH)$ then from \eq{at} the cutoff simply changes linearly with time. Such a surface, however,  does not correspond a constant $H$ surface.} cannot simply be described by $a=a_c$ (see Fig. \ref{fig3}). 

One may check that the integral of $\o_\cH$ on the  deformed surface equals to \eq{int} as follows. Since $\o_\cH=d(p_\f d\f)$, the surface integral can be reduced to a line integral along the boundary on the right.\footnote{Note that $m\f=\pm\sqrt{6}H$ does not define a boundary. They correspond to two antipodal points on $S^1$.} Denoting the the boundary of $\S_H$ as $a=a(H,\f)$,  the integral 
\bea
\int_{\S_H} \o_\cH&=&\oint p_\f d\f\\
&=&4\int_0^{\sqrt{6}H/m}a(H,\f)^3\sqrt{6H^2-m^2\f^2}\,d\f\nn
\eea
becomes independent of $H$ by \eq{a0} and \eq{ad}. 

Although $\S_H$ turns out to be  quite different than the simple $a=a_c$ surface at $H=H_*$, \eq{cutoff} can still  be a good estimate for the size of $\S_H$ with respect to the canonical measure. Nevertheless, the shape of $\S_H$ is crucial in evaluating the probability of inflation. The inflationary orbits obey $\dot{\f}\simeq0$ and $m\f\simeq 6H^2$, and  by \eq{ad} $|da/dH|$ is larger compared to non-inflationary orbits. Thus, inflationary  orbits are squeezed near the edge of the phase space pictured in Fig. \ref{fig3}, which gives the exponential suppression obtained in \cite{gt}.

In \cite{ct}, the divergence of the original measure as $a\to\infty$ is regulated by introducing a delta function, which force the measure to be concentrated on flat universes (\cite{ct} considers FRW model with the spatial  curvature).  However, this regularization does not respect the Hamiltonian flow and thus the result depends on the Hubble parameter one chooses. Here, on the other hand, we see that at a larger Hubble parameter the  regulated region deforms non-trivially to keep the measure constant.  

\subsection{Surfaces of constant $a$}

Let us now consider an alternative foliation of $\tilde{\C}$ by the surfaces of constant  scale factor. The two dimensional surface $\S:a=a_*$ is  $S^1\times R^+$, where $S^1$ is the circle in the $(\f,m\dot{\f})$ plane defined by \eq{c} and $R^+$ stands for $H$ (this is the cone $C_2$ with the origin removed). Using $\f$ and $\dot{\f}$ as coordinates in $\S$, and viewing $H$ as a function of them fixed by \eq{c}, the reduced symplectic form can be found as 
\be
\o_\cH=a_*^3\,d\dot{\f}\wedge d\f. 
\ee
As before, the integral $\int_\S\o_\cH$, which is proportional to the area of the infinite cone $C_2$,  diverges and thus $\S$ must be restricted to have a finite measure. The natural way is to impose $H\leq H_c$. Indeed,  a cutoff of the order of the Planck scale  $H_c\sim M_{Pl}$ is already necessary for the validity of the classical field equations. Before that "time", quantum gravitational effects must be taken into account. 

As in our previous discussion, the surface $\S$ evolves non-trivially under the Hamiltonian flow. From the flow equations one may find that
\be\label{hf}
\fr{dH}{da}=-\fr{\dot{\f}^2}{2aH},
\ee
where, as noted above, $H$ must be viewed as a function of $\f$ and $\dot{\f}$ defined by \eq{c}. We see that $H$ is a decreasing function of $a$, and thus the cutoff effectively reduces for $a>a_*$. However, on the evolved surface $H$ acquires different values depending on $\f$ and $\dot{\f}$ due to the explicit $\dot{\f}$ dependence in \eq{hf}. Namely, the circle $\dot{\f}^2+m^2\f^2=6H_c^2$ both shrinks and deforms along the orbits through the origin. 

\begin{figure}
\centerline{\includegraphics[width=7cm]{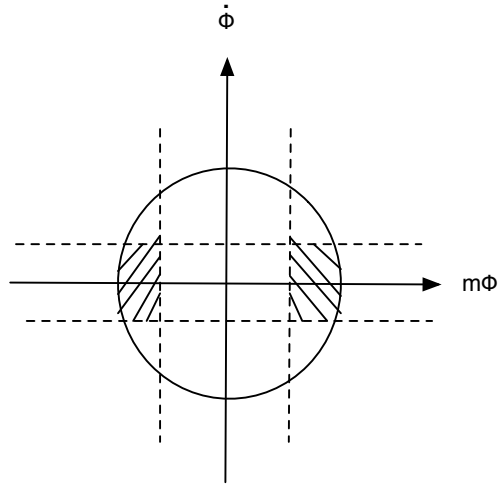}}
\caption{The finite region $\S$ restricted by$H\leq H_c$. The inflationary orbits with at least $N$ e-folds cross  the dashed area. The dotted lines are given by $\f=\pm2\sqrt{N}$ and $\dot{\f}=\pm \sqrt{2/3}m$.} 
\label{fig4}
\end{figure}

It is possible to calculate the probability of inflation in this setup as follows. First note that 
\be\label{area}
\int_\S \o_\cH=6\pi\fr{a_*^3H_c^2}{m}.
\ee 
Second, for at least $N$ e-folds of inflation the initial values must obey\footnote{These conditions can be found in various ways. The easiest root is to note that the slow-roll parameter $\e$ is related to number of e-folds $N$ by $\e\simeq 1/(2N)$ where $\e=(V'/V)^2/2=2/\f^2$. } $\f_i\geq 2\sqrt{N}$ and $\dot{\f}_i\leq\sqrt{2/3}m$ (see e.g. \cite{sm}). Therefore, the orbits which inflate more than $N$ e-folds pass through the region inside the circle $H\leq H_c$, in between the lines $\dot{\f}=\pm \sqrt{2/3}m$ obeying $|\f|\geq 2\sqrt{N}$ (see Fig. \ref{fig4}). For $H_c\gg m$, the integral of $\o_\cH$ over that region is approximately given by 
\be 
\int \o_\cH\simeq 2 a_*^3 \sqrt{\fr{2}{3}}m\left(\fr{\sqrt{6}H_c}{m}-2\sqrt{N}\right). 
\ee
Dividing this to the total area \eq{area}, one finds the probability of $N$ e-folds of inflation as\footnote{In this setup the number of e-folds $N$ turns out to have a maximum given by $3H_c^2/(2m^2)$.}  
\be
P_{N}\simeq \fr{\sqrt{2}}{3\pi\sqrt{3}}\fr{m^2}{H_c^2}\left(\fr{\sqrt{6}H_c}{m}-2\sqrt{N}\right).
\ee
We see that $P_N$  is not exponentially suppressed, but being cutoff dependent it is of the order of $m/H_c$.  On the other hand,  it becomes manifestly independent of $a_*$. For the conventional value of the scalar mass $m\sim 10^{-6}M_{Pl}$ and for $H_c\sim M_{Pl}$, one gets $P_N\sim 10^{-6}$ for $N<10^{12}$. This calculation, namely imposing initial conditions near the Plank scale cutoff  $H\simeq M_{Pl}$, is close to the standard discussion of the probability of inflation in the chaotic inflationary scenario, see e.g. \cite{klm}. 

As discussed in  \cite{sm}, the computation of probabilities is expected to depend sensitively on the regularization procedure and results of this section supports this claim. In \cite{sm}, different ways of getting  $a_c\to \infty$ limit is shown to give different results for the probability of inflation. It is clear that once a finite cutoff is introduced then there is no ambiguity in the calculation of probabilities. Thus, the main issue to be discussed is whether a proposed cutoff is physically viable or not.  On the other hand, getting two different results as a result of using two seemingly viable cutoffs weakens the validity of this construction. Note that it is possible to choose other foliations of the constrained phase space, for example one may consider $aH^n$ with $n\leq 1/3$ as the flow parameter. Presumably, one gets different results for the probability of inflation for these different choices. 
 
\subsection{The measure for non-abelian gauge fields}

There are  quite a  number of different ways of realizing inflation. In \cite{msj}, an interesting model involving non-abelian gauge fields minimally coupled to gravity with a particular $F^4$ interaction is shown to yield slow-roll inflation. Specifically, the model of \cite{msj} is based on an $SU(2)$ subgroup of a non-abelian gauge group with the gauge fields denoted by $A_\m^a$, where $a=1,2,3$ labels $SU(2)$ algebra. The action is taken as
\bea\nn
S=\fr12\int\sqrt{-g}\left[R+\fr14 F^a_{\m\n}F_a^{\m\n}\right.\hs{15}\\
\left.-\fr{\k}{192}\left(\e^{\m\n\r\s}F^a_{\m\n}F_{a\r\s}\right)^2\right],
\eea 
where $\k$ is a new dimension-full constant. To respect isotropy, the gauge field can be assumed to have the form $A_i^a=\f(t)\d^a_i$, where $i=1,2,3$ refers to the  coordinate frame in \eq{met} . With this ansatz, the action can consistently be reduced to give the following effective Lagrangian \cite{msj}
\bea\nn
L=-\fr{3a}{N}\dot{a}^2+\fr{3a}{2N}\dot{\f}^2-\fr{3Ng^2}{2a}\f^4+\fr{3\k g^2}{2Na^3}\f^4\dot{\f}^2.
\eea 
As discussed in \cite{msj}, $\f/a$ behaves like a genuine scalar  under diffeomorphisms. However, for the following analysis it is useful to keep $\f$ as the main variable since it is identical with the  gauge field components $A_\m^a$. 

The  momenta conjugate to $a$ and $\f$ can be found as
\bea
&&p_a=-6a\dot{a},\nn\\
&&p_\f=3a\dot{\f}+3\k g^2\fr{\f^4}{a^3}\dot{\f}.\nn
\eea
As usual the symplectic form is given by  $\o=dp_a\wedge da+dp_\f\wedge d\f$. Note that $p_\f$ blows up as $a\to0$, which might be thought to indicate a new divergence for the measure at $a=0$ (this turns out not to be the case as we will see below). The Hamiltonian constraint is
\be\label{gc}
2H^2=\fr{\dot{\f}^2}{a^2}+g^2\fr{\f^4}{a^4}+\k g^2\fr{\f^4\dot{\f}^2}{a^6},
\ee
which shows that $H$ is monotonic. Therefore the transverse surface can again be chosen as $\S:H=H_*$. Using $(a,\f)$ as coordinates in $\S$ and solving $p_\f$  from \eq{gc}  as
\bea\nn
p_\f=3a^2\sqrt{(2H_*^2-g^2\f^4/a^4)(1+\k g^2\f^4/a^4)},
\eea
a straightforward calculation gives
\bea\label{ing}
\int_\S \o_\cH=\int_0^{a_c}\int_{-a(2H_*^2/g^2)^{1/4}}^{a(2H_*^2/g^2)^{1/4}} \left(\fr{\del p_\f}{\del a}\right) d\f da.
\eea
Note that from $p_\f$  above, the range of $\f$ depends on the scale factor $a$, so in \eq{ing} the order of integration is crucial. Introducing a new integration variable $\f=a\psi$, \eq{ing} becomes
\be
\int_\S \o_\cH=h(H_*)\,a_c^3,
\ee
where $h$ is given by a complicated integral. Therefore, the measure only diverges for large $a$ as in the case of scalars. 

On the other hand, if the transverse surface is chosen as $\S:a=a_*$,  a cutoff for the Hubble parameter can be introduced by $H\leq H_c$. Using $(\dot{\f},\f)$ as independent coordinates, the measure then gives 
\bea\nn
\int_\S \o_\cH&=&\int\int_{H_c} (3a_*)\left[1+\k g^2\fr{\f^4}{a_*^4}\right]d\dot{\f}d\f\\
&=&a_*^3\,\tilde{h}(H_c),
\eea
where the integral is in the region constrained by \eq{gc} with $a=a_*$ and $H=H_c$, and $\tilde{h}$ is fixed by a complicated integral. 

In both cases, we see that a shift  of the transverse surface, i.e. a change of $H_*$ or $a_*$, induces a change in the cutoff $a_c$ or $H_c$, respectively. Therefore, the constrained surface, either $\S:H=H_*$ or  $\S:a=a_*$, evolves non-trivially under the Hamiltonian flow, implying a naturalness problem. 

\section{Conclusions} 

Determining how probable is inflation is an important open problem. As shown in \cite{gt}, the canonical measure offers a natural probability distribution on the set of cosmological solutions, but unfortunately it diverges \cite{hp}. The relatively recent work of \cite{gt} suggests a way of regularizing this divergence by introducing a cutoff for the scale factor, since universes larger than a critical, but yet unknown size, cannot be observationally distinguished. 

In this paper, after reviewing the basic properties of the canonical measure, we comment on the fact that the simple and the natural cutoff introduced in \cite{gt} becomes complicated when it is viewed at an earlier cosmic time. Although the measure is invariant, and thus the probabilities can be determined at any Hubble time, the variation of the cutoff  along the phase space raises concerns about the naturalness of the procedure. We also consider an alternative but physically well motivated cutoff involving the Hubble parameter and determine the probability of inflation, which turns out to be cutoff dependent but not exponentially suppressed as in \cite{gt}. 

As it is discussed in detail in a recent paper \cite{sm},  one can criticize these results on different grounds like the lack of a mechanism, which would enforce equilibration on cosmological scales and thus imply a flat probability distribution in the phase space. However, in the absence of a fundamental theory, which is expected to predict the initial state of the universe in some way, the canonical measure offers a viable way of addressing the issue. Using the canonical measure, one can at least try understand if our universe is typical or not under the assumption that all universes in the multiverse is "equally" probable. Therefore, it is important to find a satisfactory way of making the measure finite, presumably formulating it an a fundamental theory, and determine the probability of inflation under certain assumptions. In that way one would get a better idea if inflation is common/natural or not.

\end{document}